\documentclass[twocolumn,showpacs,preprintnumbers,amsmath,amssymb]{revtex4}
\usepackage{graphicx}
\begin{document}

\title{Incompressible excitonic superfluid of ultracold Bose atoms
in an optical lattice: a new superfluid phase in the one-component
Bose-Hubbard model}

\author{Yue Yu }
\affiliation{Institute of Theoretical Physics, Chinese Academy of
Sciences, P.O. Box 2735, Beijing 100080, China}
\date{\today}
\begin{abstract}
We predict that a new superfluid phase, the incompressible
excitonic superfluid (IESF), in the phase diagram of ultracold
Bose atoms in $d>1$ dimensional optical lattices, which is caused
by the spontaneous breaking of the symmetry of translation of the
lattice. Within mean field theory, the critical temperature of the
phase transition from this IESF to the normal fluid (NF) is
calculated and the triple-critical point of the three phases is
determined. We also investigate both configuration and gauge field
fluctuations and show the IESF state is stable against these
fluctuations. We expect this IESF phase can be experimentally
observed by loading cold Bose atoms into a two-dimensional lattice
where the atom filling fraction deviates slightly from exact
commensurations. The signatures distinguishing this IESF from the
common atom superfluid (ASF) are that (i) the critical temperature
of the IEST/NF transition is independent of interaction, unlike
the ASF/NF transition; (ii) the IESF is incompressible while the
ASF is compressible.
\end{abstract}


\maketitle

Ultracold atoms in optical lattices offer new opportunities to
study strongly correlated phenomena in a highly controllable
environment\cite{1,2,3,kohl}. A quantum phase transition, the
superfluid/Mott-insulator transition, was demonstrated using
$^{87}$Rb atoms \cite{2,kohl}. Strongly correlated phenomena for
boson systems may be studied theoretically by the Bose-Hubbard
model \cite{BH,jaks}. Most studies focused on the quantum phase
transition from the superfluid to the Mott insulator for a
commensurate lattice at zero temperature.

Recently, the phase diagram of the ultracold Bose atoms has been
investigated by Dickerscheid et al \cite{dick} by a slave boson
approach and greatly improved by a slave fermion approach
\cite{yu}. It was claimed that the Mott insulator phase crossovers
to the normal fluid (NF) in a finite temperature for a
commensurate lattice. For an incommensurate lattice, it was
believed that below a critical temperature, the atoms are in the
superfluid phase even in the strong interaction limit if the
normal state is thought as a NF. To our knowledge, the phase
structure of the normal state has not been thoroughly analyzed.

We define the 'normal' state of the one-component ultracold bose
atoms in $d>1$ optical lattices by the vanishing of the order
parameter of the common atom superfluid(ASF).  We find that there
is an incompressible (atom-hole) excitonic superfluid (IESF)
phase, due to the spontaneous breaking of the symmetry of
translation of the lattice. For the commensurate filling fraction,
this IESF phase is right above the zero temperature Mott
insulator. Moreover, for an integer filling fraction, creating an
atom-hole pair is accompanied by a double occupation. Thus, the
IESF phase is enshrouded by the crossover from the Mott insulator
to NL. To observe this IESF phase, the atom filling fraction of
the lattice has to be incommensurate. As we will see that the
critical temperature of the NF/IESF transition is very low. If the
filling fraction deviates from the commensuration, say one percent
from the one atom per site, the critical temperature of the NF/ASF
transition will be higher than the critical temperature of the
NF/IESF transition for a quite strong interacting strength. Thus,
to observe such an IESF state, the filling fraction has to be not
exact but very close to an integer. The justification to
distinguish this IESF from the ASF is that the critical
temperature of the IESF/NF transition is not interacting-dependent
while the critical temperature of the ASF/NF transition decreases
as the interaction strength increases. On the other hand, the IESF
is incompressible while the ASF is compressible.

We investigate the ultracold atoms by the Bose-Hubbard model
\cite{BH,jaks} with Hamiltonian $ H=-t\sum_{\langle
ij\rangle}a^\dagger_ia_j-\mu\sum n_i+\frac{U}2\sum_i
n_i(n_i-1)+V_{trap}$. Here $a_i^\dagger$ is a Bose atom creating
operator on site $i$ and the symbol $\langle ij\rangle$ denotes
the sum over all nearest neighbor sites. $\mu$ is the chemical
potential. $t$ is the hopping amplitude and $U$ the on-site
interaction, which are determined by the optical lattice
parameters and the $s$-wave scattering length of the atoms. We
study $d>1$ optical lattices only in this work. We first consider
a homogeneous case with $V_{trap}=0$. In order to explore the
finite temperature behavior, we decompose the boson operator by
the slave fermions, $
a^\dagger_i=\sum_{\alpha=0}^\infty\sqrt{\alpha+1} c_{\alpha+1,
i}^\dagger c_{\alpha,i}, $ where the slave fermion operators
$c_{\alpha,i}$ obey
$\{c_{\alpha,i},c^\dagger_{\beta,j}\}=\delta_{\alpha\beta}\delta_{ij}$
. As the auxiliary particles, they have to obey the constraint $
\sum_\alpha n^\alpha_i=\sum_\alpha
c^\dagger_{\alpha,i}c_{\alpha,i}=1$ on each site. In the slave
fermion language, the partition function of the system reads
 $Z={\rm Tr} e^{-\beta H}=\int Dc_\alpha Dc^\dagger_\alpha
D\lambda ~e^{-S_E}$ where \cite{yu}
\begin{eqnarray}
&&S_E[\bar c_\alpha, c_\alpha,\lambda]=\int_0^{1/T} d\tau
\biggl\{\sum_i\sum_{\alpha}
c^\dagger_{\alpha,i}[\partial_\tau-\mu\nonumber\\
&&~+\frac{U}2\alpha(\alpha-1)
+i\lambda_i]c_{\alpha,i}-i\sum_i\lambda_i \\
&&~~-t\sum_{\langle
ij\rangle}\sum_{\alpha\beta}\sqrt{\alpha+1}\sqrt{\beta+1}
c^\dagger_{\alpha+1,i}c_{\alpha,i}c^\dagger_{\beta,j}
c_{\beta+1,j}\biggr\},\nonumber
\end{eqnarray}
where $\lambda_i$ is a Lagrange multiplier  field and $n$ is the
atom filling fraction of the lattice.

To study the ASF/NL transition, we decouple the four slave fermion
term by introducing a Hubbard-Stratonovich field $\Phi_i$,  which
is a bosonic field and may be identified as the order parameter of
the common atom superfluid. In our recent work, we have given the
details to describe this phase transition by the slave fermion
approach \cite{yu}. We find that using a finite type slave fermion
approximation, say the maximal $\alpha_M=6$, the well-known
special points in the phase diagram may be well reproduced
\cite{lly}. For example, for the filling fraction $n=1$, the
non-interacting Bose gas in a three-dimensional lattice has the
critical temperature $T_c\approx 7.08t$ while our
 calculation for $\alpha_M=6$ gives $T_c\approx 7.01t$; the zero
temperature critical interacting strength from the superfluid to
Mott insulator is $U_c/zt\approx 5.82$  in the mean field theory
\cite{mean} while our result is $U_c/zt\approx 5.9$.

We now focus on the 'normal' state in which
$\langle\Phi_i\rangle=0$. We decompose the four slave fermion term
by auxiliary Hubbard-Stratonovich fields
$\hat\chi_{\alpha\beta,ij}$ and $\hat\eta_{\alpha\beta,ij}$,
namely, the Lagrangian of the system is rewritten as
\begin{eqnarray}
L&=&\sum_i\sum_\alpha c^\dagger_{\alpha
i}[\partial_\tau-\alpha\mu+\frac{U}2
\alpha(\alpha-1)-i\lambda_i]c_{\alpha
i}\nonumber\\&+&i\sum_i\lambda_i+\frac{1}2\sum_{\langle
ij\rangle}\sum_{\alpha\beta}t_{\alpha\beta}
[\hat\chi^\dagger_{\alpha\beta,ij}\hat\chi_{\alpha+1\beta+1,ij}
\nonumber\\
&-&\hat\eta^\dagger_{\alpha\beta,ij}(\hat\chi_{\alpha+1\beta+1,ij}
-c^\dagger_{\alpha+1,i}c_{\beta+1,j}) +h.c.],
\end{eqnarray}
where $t_{\alpha\beta}=t\sqrt{\alpha+1}\sqrt{\beta+1}$.
Integrating over $\hat\eta$ and $\hat\chi$, the above Lagrangian
restores the original one. The auxiliary fields
$\hat\chi_{\alpha\beta,ij}$, $\hat\eta_{\alpha\beta,ij}$ and
$\lambda_i$ may be rewritten as
$\hat\chi_{\alpha\beta,ij}=\chi_{\alpha\beta,ij}e^{i{\cal
A}_{ij}}$,
$\hat\eta_{\alpha\beta,ij}=\eta_{\alpha\beta,ij}e^{i{\cal
A}_{ij}}$  and $\lambda_i=\lambda+{\cal A}_{i0}$. Consider the
mean field state,
$\hat\chi_{\alpha\beta,ij}\approx\chi_{\alpha,ij}\delta_{\alpha\beta}$,
$\hat\eta_{\alpha\beta,ij}\approx\eta_{\alpha,ij}\delta_{\alpha\beta}$,
and $\lambda_i\approx\lambda$. The mean field Lagrangian reads
\begin{eqnarray}
L_{MF}&=&\sum_{i,\alpha} c^\dagger_{\alpha i}M_{\alpha}c_{\alpha
i} -\sum_{\langle ij\rangle}\sum_{\alpha}\frac{1}2[t_{\alpha}
(\eta^\dagger_{\alpha,ij}-\chi^\dagger_{\alpha,ij})\chi_{\alpha+1,ij}\nonumber\\
&-&(t_{\alpha-1}\eta_{\alpha-1,ij}^\dagger
+t_\alpha\eta_{\alpha+1,ji})c^\dagger_{\alpha i}c_{\alpha
j}+h.c.],\label{mfl}
\end{eqnarray}
where $M_{\alpha}=\partial_\tau-\alpha\mu+\frac{U}2
\alpha(\alpha-1)-i\lambda$.  The free single slave fermion Green's
functions can be read out from (\ref{mfl}),i.e., $\hat
D_\alpha(x,\tau)=T\sum_n e^{i\omega_n\tau}\hat D_\alpha(\omega_n)$
with the matrix
\begin{eqnarray}
\hat D_\alpha(\omega_n)&=&[(i\omega_n+\frac{U}2\alpha(\alpha-1)
-\mu+i\lambda)\delta_{ij}\nonumber\\&+&t_{\alpha-1}\eta_{\alpha-1,ij}^\dagger
+t_\alpha\eta_{\alpha+1,ji}]^{-1}.
\end{eqnarray}
The mean field equations are given by
$\chi_{\alpha,ij}=\eta_{\alpha,ij}$ and $\eta_{\alpha,ij}=\langle
c_{\alpha i}^\dagger c_{\alpha j}\rangle=T\sum_n
D_{ij,\alpha}(\omega_n)$. Near the critical temperature, these
mean field equations have the following solutions
\begin{eqnarray}
\eta_{\alpha,ij}T^{(\alpha)}_c=\frac{t_{\alpha-1}\eta_{\alpha-1,ij}^\dagger
+t_\alpha\eta_{\alpha+1,ij}^\dagger}
{(e^{\beta^{(\alpha)}_c(\frac{U}2\alpha(\alpha-1)-\alpha
\mu-i\lambda)}+1)^2}.
\end{eqnarray}
For $\alpha\geq 2$, the critical temperatures vanish,
$T^{(\alpha)}_c=0$, while
\begin{eqnarray}
T^{(0)}_c=\frac{\eta^\dagger_{1}}{4\eta_{0}}t,~~~~~T^{(1)}_c\approx
\frac{\eta^\dagger_{0}}{\eta_{1}}t,
\end{eqnarray}
if $\beta_c\mu\gg 1$. The coupling between $\eta_0$ and $\eta_1$
means $\eta_1=2\eta^\dagger_0$ and $T_c^{(0)}=T_c^{(1)}=T_c=t/2$
\cite{note}. Changing the variable $\Delta_0=-i\eta_0/t$ and using
$\eta_1=2\eta^\dagger_0$, the mean field free energy near the
critical temperature may be expanded by \cite{amil}
\begin{eqnarray}
 F&=&\sum_{\langle ij\rangle}
(2t^{-1}\Delta_{0,ij}^2
-T^{-1}|\Delta_{0,ij}|^2)\nonumber\\
&+&\frac{1}{24 T_c^3}\sum_{\langle ijkl\rangle}
\Delta_{0,ij}\Delta_{0,jk}\Delta_{0,kl}\Delta_{0,li}.\label{fe}
\end{eqnarray}
The first term in (\ref{fe}) implies only a real $\Delta_{0,ij}$
minimizes the free energy, which is given by
\begin{eqnarray}
F=-3 t\tau^2 S_2^2/S_4,
\end{eqnarray}
where $\tau=\frac{T_c-T}T$; $S_2$ and $S_4$ are the numbers of the
non-zero terms of the first and second summations in (\ref{fe}).
Due to the filling fraction very close to $n=1$, we consider all
holes in the lattice are isolated. There are two solutions. One is
the equal-bond state. Namely, for a hole at site $i$,
$\Delta_{0,i,i\pm\tau_x}=-\Delta_{0,i,i\pm\tau_y}=\Delta_0$ for
all nearest neighbor sites $j$ and all others are zero. Another
solution is an atom-hole exciton state and
$\Delta_{0,ij}=\Delta_0$ is not zero only if a link $(ij)$ is
occupied by an exciton, {\it which breaks the symmetry of
translation of the lattice in the link direction}. The free
energies per site for the equal-bond and exciton states are
degenerate, e.g., for a two-dimensional square lattice with a low
hole density,
\begin{eqnarray}
f_{eb}=f_e=-6n^0t\tau^2,
\end{eqnarray}
where $n^0$ is the hole density. There is, however, a
configuration fluctuation to the equal-bond state. When two holes
are close so that they are in the diagonal line of a plaquette,
the equal-bond state will gain the free energy an amount
$\frac{4}3t\tau^2$ while the free  energy of the exciton state
does not change. Multi-hole configurations further raise the
energy of the equal-bond state. On the other hand, due to the
gauge fluctuation as we shall see the equal-bond state may not be
stable.

To study the gauge fluctuations, we first calculate the dispersion
relations of the slave fermions for these two mean field states.
For a two-dimensional square lattice, we divide the lattice into
two sublattices, even and odd. The slave fermion operators
$c_{\alpha,i}$ is denoted by either $e_{\alpha,i}$ or
$d_{\alpha,i}$, corresponding to $i\in$ even or odd, respectively.
Using the Fourier components
$e_{\alpha,k}=\frac{1}{\sqrt{N/2}}\sum_{i\in even} e^{-ik\cdot
R_i}e_{\alpha,i}$ and $d_{\alpha,k}=\frac{1}{\sqrt{N/2}}\sum_{i\in
odd} e^{-ik\cdot R_i}d_{\alpha,i}$, the hopping term of the
Hamiltonian in the mean field theory may be diagonalized
\begin{eqnarray}
H_t=\sum_{\alpha=0,1;k}|\Delta^{(\alpha)}(k)|
(\zeta_{\alpha,k}^\dagger\zeta_{\alpha,k}- \xi_{\alpha,k}^\dagger
\xi_{\alpha,k}),
\end{eqnarray}
where $\Delta^{(0)}(k)=\Delta_1(k)$ and
$\Delta^{(1)}(k)=\Delta_0(k)$.
$\Delta_{\alpha}(k)=\Delta_{\alpha,1}e^{ik_x}
-\Delta_{\alpha,2}e^{-ik_y}+\Delta_{\alpha,3}e^{-ik_x}
-\Delta_{\alpha,4}e^{ik_y}$ where 1,2,3 and 4 denote four adjacent
sites around a hole. The diagonalized operators are defined by
\begin{eqnarray}
&&e_{\alpha,k}=\sqrt{\frac{1}2}(\xi_{\alpha,k}+\frac{i\Delta^*_{\alpha,k}}
{|\Delta_{\alpha,k}|}\zeta_{\alpha,k}),\nonumber\\
&&d_{\alpha,k}=\sqrt{\frac{1}2}(-\frac{i\Delta^*_{\alpha,k}}
{|\Delta_{\alpha,k}|}\xi_{\alpha,k}+\zeta_{\alpha,k}).
\end{eqnarray}
For the equal-bond state, the slave fermion dispersions are
$\epsilon_{eb,\alpha}(k)=\pm2|\Delta_\alpha|(\cos k_x+\cos k_y)$,
which have gapless excitations. For the exciton, the slave
quasi-fermions are dispersionless with
$\epsilon_{d,\alpha}(k)=\pm|\Delta_\alpha|$, which means that
exciting a slave quasi-fermion has an energy gap
$|\Delta_\alpha|$. However, the slave fermions are auxiliary
particles and the single slave fermion Green's function is not
gauge invariant. To see the real quasiparticle excitation, one has
to calculate a gauge invariant Green's function. The single-atom
thermodynamic Green's function is gauge invariant, which reads
\begin{eqnarray}
&&\langle T(a(x,t)a^\dagger(0,0))\rangle=\sum_{\alpha\beta}
\sqrt{\alpha+1}\sqrt{\beta+1}\nonumber\\ &&\langle
T(c^\dagger_{\alpha}(x,t)c_{\alpha+1}(x,t)
c^\dagger_{\beta+1}(0,0)c_{\beta}(0,0))\rangle.
\end{eqnarray}
If we consider the lowest lying excitation, only relevant
propagating processes are those which do not raise an additional
on-site energy $U$. That is, at $t=0$, positions $x$ and 0 have
the occupation numbers $\alpha$ and $\alpha+1$, respectively, and
after a time $t$, one atom propagates from 0 to $x$, the
occupation numbers become $\alpha+1$ at $x$ and $\alpha$ at 0. The
approximation we are using yields the following factorization of
the four slave fermion Green's function, $\sum_{\alpha\beta}
\sqrt{\alpha+1}\sqrt{\beta+1}\langle
T(c^\dagger_{\alpha}(x,t)c_{\alpha+1}(x,\tau)
c^\dagger_{\beta+1}(0,0)c_{\beta}(0,0))\rangle \approx\sum_\alpha
(\alpha+1)\langle T(c^\dagger_{\alpha}(x,t)c_{\alpha}(0,0))
\rangle\langle T(c_{\alpha+1}(x,t)c^\dagger_{\alpha+1}(0,0))
\rangle$. The corresponding retarded Green's function is given by
\begin{eqnarray}
G^R(\omega)\propto \sum_{\alpha}
\frac{(\alpha+1)(n^\alpha-n^{\alpha+1})}{\omega-\alpha
U+|\Delta^{(\alpha+1)}|-|\Delta^{(\alpha)}|+\mu+i0^+}.
\end{eqnarray}
where the order parameters $\Delta^{(0)}=\Delta_1$,
$\Delta^{(1)}=\Delta_0$ and vanish for others.

Thus, the lowest energy quasiparticle excitation at finite
temperature is
\begin{eqnarray}
\varepsilon(k)=|\Delta_1(k)|-|\Delta_0(k)|=|\Delta_0(k)|.
\end{eqnarray}
This implies the low lying excitations of the exciton state has a
gap $|\Delta_0|$, which is equal to $t$ at $T=0$ , while it is
gapless for the equal-bond state. The second level excitation
spends an energy $U-|\Delta_1(k)|$.

\begin{figure}
\begin{center}
\includegraphics[width=8.5cm]{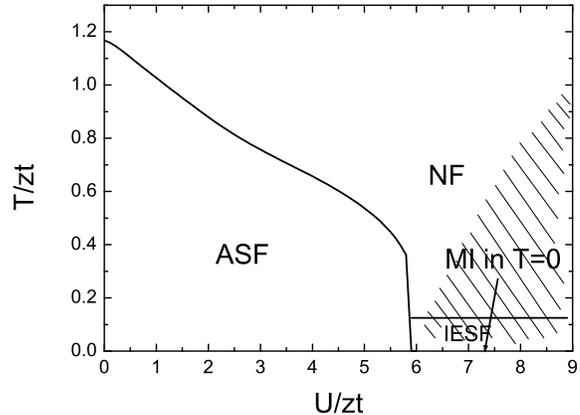}
\end{center}
 \caption{ The phase diagram for 2d for n=1. ASF, NF, IESF and MI are
 standing for the atom superfluid, normal fluid, incompressible excitonic
 superfluid and
 Mott insulator, respectively. The shadowed area is the crossover
 regime from MI to NF.
   }\label{fig1}
\end{figure}

\begin{figure}
\begin{center}
\includegraphics[width=8.5cm]{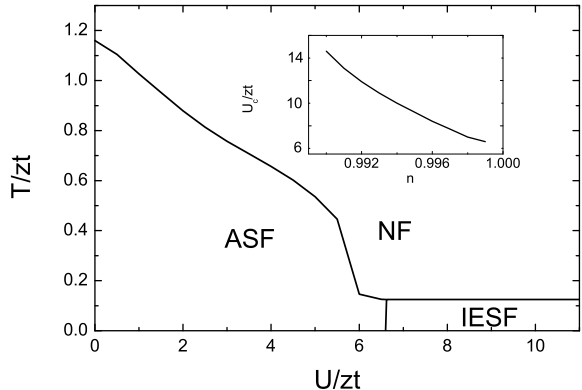}
\end{center}
 \caption{ the phase diagram for 2d for n=0.999. The inset is the
 value of $U_c$ in the triple-critical varying as the atom
  filling fraction.
   }\label{fig2}
\end{figure}

We now discuss the gauge fluctuations. The gauge fluctuations come
from the gauge field ${\cal A}_{ij}$ and ${\cal A}_{0i}$. For the
equal-bond state, the gauge fluctuations may be more serious than
in the $t$-$J$ model because we do not have a large $N$ limit.
However, the existence of the gap in the exciton state strongly
suppresses the gauge fluctuations. Integral over the slave fermion
field, one can get the effective action of the gauge field
\begin{eqnarray}
&&S[a]=\frac{T}2\int
d^2ka_\mu(k,\omega)a_\nu(k,\omega)\Pi_{\mu\nu}(k,\omega),\nonumber\\
&&\Pi_{\mu\nu}=\sum_\alpha\int
d^2p\biggl[\frac{n^\alpha(p+k/2)-n^\alpha(p-k/2)}
{\omega-[\epsilon_\alpha(p+k/2)-\epsilon_\alpha(p-k/2)]}\nonumber\\
&&\times \frac{\partial \epsilon_\alpha}{\partial
p_\mu}\frac{\partial \epsilon_\alpha}{\partial p_\nu}
+\frac{\partial^2\epsilon_\alpha}{\partial p_\mu\partial
p_\nu}n(p)\biggr],
\end{eqnarray}
where $a_\mu(k,\omega)$ is the Fourier component of the continuum
limit of the gauge field.  The dispersionless of the exciton state
means $\Pi_{\mu\nu}=0$ which suppresses the gauge fluctuation. The
physical meaning of $\Pi_{\mu\nu}$ is the atom density-density and
current-current correlation functions. The vanishing of the
density-density correlation function yields the incompressibility
of the state.

For the equal-bond state, the gauge field propagator
$\Pi_{\mu\nu}^{-1}=\langle a_\mu(q)a_\nu(-q)\rangle =
(\delta_{\mu\nu}-q_\mu q_\nu/q^2)D_T (\vec q)$ with $D_T (\vec q)
\simeq [i\omega/q-\chi_d q^2]^{-1}$  for $\omega<q$  \cite{nl}.
The gauge fluctuation may renormalize the hopping amplitude $t$ to
$\tilde t=t \langle e^{i{\cal A}_{ij}}\rangle=te^{-\langle {\cal
A}_{ij}^2\rangle}$ and the critical temperature is reduced to
$\tilde T_c=\tilde t/2$.

To sum up, both configuration and gauge fluctuations destroy the
equal-bond state while the exciton state is safe.  The excitons
form an incompressible fluid. Furthermore, because
$\Delta_1=2\Delta_0\ne 0$, the order parameter $\langle
a^\dagger_ia_j\rangle$ of the exciton condensate is not zero. This
means the excitons are condensed when $T<T_c$. This is an
incompressible excitonic superfluid (IESF).

The phase diagrams of the system now may be depicted according to
the above discussions. Fig.\ref{fig1} is the phase diagram for the
filling fraction $n=1$ for two-dimensions. There is a
triple-critical point with $U_c/zt\approx 5.8$ and $T_c/zt=1/2z$.
The IESF is right above the Mott insulator. Therefore, this IESF
may not be observed in the commensurate filling fraction because
the IESF is still in the range of the practical Mott insulator.
The crossover from the Mott insulator to the normal fluid
enshrouds this IESF phase. To observe the IESF phase, one should
work in an incommensurate filling fraction. However, the critical
temperature $T_c =t/2$ is so low that the $U_c/zt\sim 15$ for
$n=0.99$ in two-dimensions. In the inset of Fig. \ref{fig2}, we
show the critical $U_c$ for $0.99\leq n\leq 0.999$ . Thus, the
IESF can only be observed when the filling fraction deviates
slightly from the commensuration. In Fig. \ref{fig2}, we show the
phase diagram for $n=0.999$ for two dimensions. For three
dimensions, the triple-critical point is $U_c/zt\sim 16$ for
$n=0.999$, much stronger than that in two dimensions.

We now briefly discuss the inhomogeneous case with a trap
potential $V_{trap}=V\sum_i r^2_i n_i$ where $r_i$ is the distance
of an atom from the center of the trap. In the regime of $U$ we
are concerning, there is a central Mott plateau which is
surrounded by a superfluid ring \cite{qm}. Our result points out
that there, in fact, are two superfluid rings. The IESF ring right
surrounds the Mott plateau and the ASF ring is outside of the IESF
ring. As $U$ is enhanced, while the radius of the Mott plateau
almost will not vary, the radius of the zero compressibility
valley will be enlarged because the IESF ring becomes wider. This
requires an examination by a quantum Monte Carlo calculation,
which will be reported elsewhere because it will take a long
computing time period and more space to describe.

In conclusion, we predicted a new superfluid phase, the
incompressible excitonic superfluid, in the one-component
Bose-Hubbard model for $d>1$ if the on-site interaction is large
enough. This may be observed when the atom filling fraction
deviates slightly from a commensurate one. Two characteristics of
the IESF may be distinguished from the atom superfluid: (i) the
incompressibility of the IESF with a gap $\sim0.1\mu$K at $T=0$
for $^{87}$Rb (The Mott gap $>2.5\mu$K.); and  (ii) the
$U$-independence of the critical temperature $T_c=t/2\sim
0.05\mu$K. We calculated the free energy in a mean field theory
and analyzed the configuration and quantum fluctuations. We intend
to examine this IESF phase by a quantum Monte Carlo calculation.

The author is grateful to Y. S. Wu, T. Xiang, L. You and L. Yu for
useful discussions. This work was supported in part by Chinese
National Natural Sciences Foundation.


\begin{references}
\bibitem{note} According to \cite{yu}, the critical temperature of the
NF/ASF transition may be lower than $t/2$ only when $U/zt>5.5$. We
also found $\mu\sim U/2$. Thus, $\mu/zt\approx U/2zt>5.5/2$, which
gives $\beta_c\mu=5.5z\gg 1$.
\end{references}
\end{document}